\documentclass[aps,prl,floatfix,twocolumn,showpacs,reprint,superscriptaddress]{revtex4-1}
\usepackage{amsfonts}
\usepackage{mathrsfs}
\usepackage{amsmath}
\usepackage{color}
\usepackage{graphicx}
\usepackage{bm}
\usepackage{amssymb}
\usepackage{xspace}
\usepackage{epstopdf}
\usepackage{dcolumn}
\usepackage{longtable}
\usepackage{multirow}
\usepackage[colorlinks=true, letterpaper=true, pdfstartview=FitV, linkcolor=blue, citecolor=blue, urlcolor=blue]{hyperref}

\makeatletter

\newcommand{\Rmnum}[1]{\expandafter\@slowromancap\romannumeral #1@}
\makeatother
\begin{document}
\title{Multivariable Scaling for the Anomalous Hall Effect}

\author{Dazhi Hou}%
\author{Gang Su}%
\author{Yuan Tian}%
\author{Xiaofeng Jin}%
\email{xfjin@fudan.edu.cn }
\affiliation{State Key Laboratory of Surface Physics and Department of Physics, Fudan University, Shanghai 200433, China; Collaborative Innovation Center of Advanced Microstructures, Fudan University, Shanghai, 210433}


\author{Shengyuan A. Yang}
\email{shengyuan\_yang@sutd.edu.sg }
\affiliation{Research Laboratory for Quantum Materials and EPD Pillar, Singapore University of Technology and Design, Singapore 487372, Singapore}
\author{Qian Niu}
\email{niu@physics.utexas.edu }
\affiliation{Department of Physics, The University of Texas at Austin, Austin, Texas 78712, USA}
\affiliation{School of Physics, International Center for Quantum Materials and Collaborative Innovation Center of Quantum Matter, Peking University, Beijing 100871, China}

\begin{abstract}
We derive a general scaling relation for the anomalous Hall effect in ferromagnetic metals involving multiple competing scattering mechanisms,
described by a quadratic hypersurface in the space spanned by the partial resistivities.  We also present experimental findings, which show strong deviation from
previously found scaling forms when different scattering mechanism compete in strength but can be nicely explained by our theory.
\end{abstract}
\maketitle

Despite recent success in recognizing the importance of intrinsic Berry
curvature in the anomalous Hall effect (AHE)~\cite{KLfirstpaper,PhysRevB.59.14915,Jungwirth2002,onoda2002,Fang03102003,Yao2004,TianYuan},
there remains a basic conflict between theories and experiments on the
scaling behavior of the anomalous Hall resistivity~\cite{nagareview}. Working under the
assumption of a single type of scatterers, theories invariably show that
the anomalous Hall resistivity is a sum of linear and quadratic terms in
the longitudinal resistivity, with the former identified as from skew
scattering~\cite{smit} and with the latter as from both intrinsic and side jump
mechanisms~\cite{berger}. However, many experiments do not follow such a simple scaling relationship.

Recent experimental~\cite{TianYuan,Hou2012,JHzhaoPRB} and theoretical studies~\cite{Nagaosa_re,Yang2011} indicate
that a major cause of the breakdown of the simple scaling
is the presence of more than one source of scatterers.
By working with molecular beam epitaxial films, Tian \emph{et al.}~\cite{TianYuan} were able to limit
scattering of electrons to two sources, one by interface
roughness and one by phonons, with independent control on their
strengths through the film thickness and sample temperature.
By taking into account the distinction of scaling parameters with
respect to different scatterings, a new phenomenological scaling form
were found which fits the experimental data much better. However,  its theoretical
derivation has been lacking so far.

In this Letter, we derive the general scaling form of the anomalous Hall
resistivity as a function of multiple competing scattering mechanisms. This is also motivated by the new experimental data to be
presented in this work, which deviates from our earlier results
in the regimes when two sources of scatterings compete in strength. We find that the anomalous Hall resistivity is generally a quadratic surface
passing through origin in the space of partial longitudinal resistivities. This reduces to
the simple one-variable scaling along each axis, but contains an extra
parameter that describes competition between the two scatterings.

The general scaling relation obtained from our theory with multiple competing scatterings can be written in the following form
\begin{equation}\label{rhoAH}
\rho_\text{AH}=c\rho_{xx}^2+\sum_i c_i\rho_i \rho_{xx}+\sum_{ij}c_{ij}\rho_i\rho_j+\sum_{i\in\text{S}}\alpha_i\rho_i,
\end{equation}
where the Hall resistivity $\rho_\text{AH}$ is expressed using the partial longitudinal resistivities  $\rho_{i(j)}$ as scaling variables which are changed through the disorder concentrations of each scattering source, $i$ and $j$ label different scattering sources, the coefficient $c$ encodes the intrinsic Berry curvature contribution, and coefficients $c_i$, $c_{ij}$, and $\alpha_i$ remain constant when disorder concentrations are changed (they do depend on the source of scattering as indicated by the indices), and $i\in\text{S}$ indicates that only static disorder scatterings are included in the summation. With Matthiessen's rule $\rho_{xx}=\sum_i\rho_i$, it can be seen that geometrically $\rho_\text{AH}$ spans a quadratic hypersurface in the space of $\rho_i$'s. Along each $\rho_i$-axis, the scaling takes the form of a parabolic curve $\alpha_i\rho_i+(c+c_i+c_{ii})\rho_i^2$ passing through the origin, which is reminiscent of the old simple scaling~\cite{nagareview} (which applies only for a single type of scattering) with the linear term ($\alpha_i$) from skew scattering and the quadratic term from both intrinsic ($c$) and side jump $(c_i+c_{ii})$ mechanisms. Furthermore, cross terms of $\rho_i\rho_j$ generally exist, manifesting the competition between different scatterings.

{\color{blue}{\it Sketch of theoretical derivation.}}---In the following, we provide a justification of our general scaling formula from microscopic theory. Although the different mechanisms have been envisaged from the early theories of AHE, theories which could systematically account for all the mechanisms were developed relatively recently~\cite{nagareview}. It has been shown that the different approaches are equivalent in the good metal regime with $\varepsilon_F\tau\gg 1$ ($\varepsilon_F$ is the Fermi energy and $\tau$ is the scattering time) which is what we are are concerned with here.

Theoretically, it is more convenient to calculate conductivity which, in Kubo formalism, is related to the equilibrium current-current correlation function and can be casted into a familiar form of the Kubo-Streda formula~\cite{stre1982}. There are two terms for the Hall conductivity $\sigma_\text{AH}$, one of which ($\sigma_\text{AH}^\text{I}$) is due to states near the Fermi surface which contains all the important scattering effects, and the other term $\sigma_\text{AH}^\text{II}$ involves all the occupied states below the Fermi level and only contributes to the intrinsic part. Hence to assess the scattering effects, one only needs to focus on
$\sigma_\text{AH}^\text{I}=\frac{e^2}{2\pi A}\text{Tr}\langle \hat{v}_x \hat{G}^R(\varepsilon_F) \hat{v}_y \hat{G}^A(\varepsilon_F)\rangle_c
$. Here $A$ is the system size, $\hat{v}_{x}$ and $\hat{v}_{y}$ are the velocity operators, $\hat{G}^R$ and $\hat{G}^A$ are the retarded and advanced Green functions respectively, and $\langle\cdots\rangle_c$ indicates a disorder configuration average. The usual practice is to treat scattering as perturbation and do expansion in the small parameter $(\varepsilon_F\tau)^{-1}$.  For simple models with single type of scattering, one can identify relevant diagrams corresponding to intrinsic, side jump, and skew scattering contributions, and group the terms according to their dependence on disorder density $n_i$, as discussed in detail by Sinitsyn \emph{et al.}~\cite{sini2007} and reviewed by Nagaosa \emph{et al.}~\cite{nagareview}.
Because both skew and side jump contributions depend on the detailed
form or composition of scattering potentials, the variation of which along the change of experimental parameters can naturally cause the
complication, especially when more than one source of scatterings are competing. However, since our focus is on the scaling relation, it is sufficient to extract the general structure and to link $\sigma_\text{AH}$ to the longitudinal transport coefficient (which in Kubo formalism takes similar form as $\sigma_\text{AH}^\text{I}$ by replacing $\hat{v}_{y}$ with $\hat{v}_{x}$) without performing detailed calculations. Our analysis below applies to general multiband systems with multiple sources of scatterings.

\begin{figure}
\includegraphics[width=7.8cm]{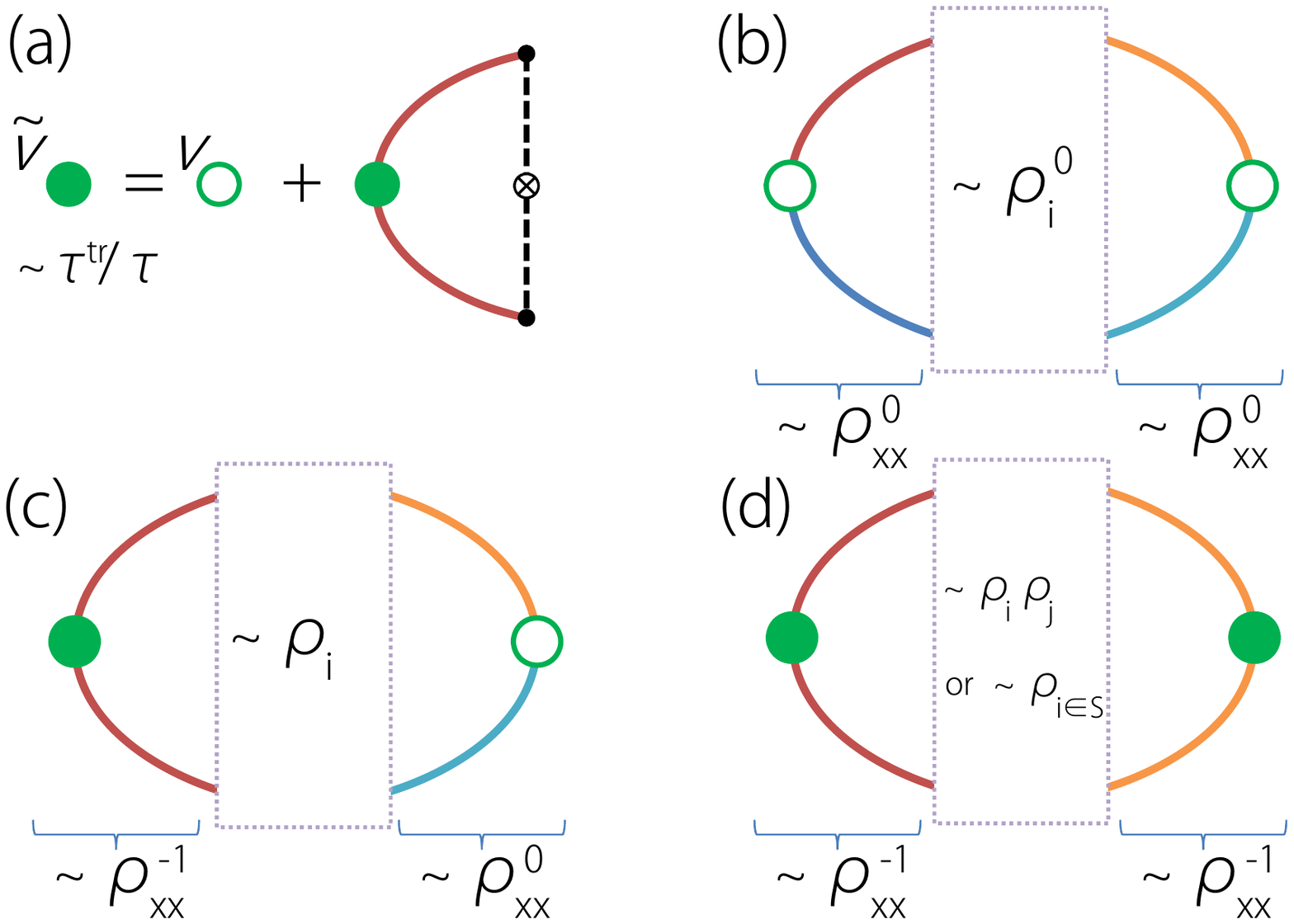}
\caption{(a) Correction of velocity vertex by a ladder diagram. The dressed vertex $\tilde{v}$ contains factor $\tau^\text{tr}/\tau$. (b-d) correspond to our defined groups 1 to 3 respectively. The colors of two Green function lines at each vertex indicate whether they belong to the same band. The dependence of leading order terms of each part on longitudinal resistivities is indicated.}
\label{fig1}
\end{figure}

For longitudinal transport in metals, it is well-known that the scattering has two important effects~\cite{mahan}: first, it introduces a finite lifetime for quasiparticles through self-energy correction to the Green functions; second, it corrects the velocity vertex, making $\sigma_{xx}\propto \tau^\text{tr}$, where $\tau^\text{tr}$ is the transport scattering time. In the so-called ladder approximation, the velocity vertex is dressed by a ladder diagram as shown in Fig.\ref{fig1}(a). Note that only intraband velocity vertex needs to be dressed, because the product involved, $[\hat{G}^R]_{n\bm k}[\hat{G}^A]_{m\bm k}$, is large ($\propto \tau$) to compensate the small factor from adding a scattering line ($\propto 1/\tau$) only when the band indices $m$ and $n$ are equal; while for $n\neq m$, the product can be at most regarded as a factor of $\tau^0$. The renormalized velocity vertex $\tilde{v}$ acquires a factor of $\tau^\text{tr}/\tau$, and when combined with the two Green function legs, one has $[\hat{G}^R]_{n\bm k} \tilde{v}_{nn}[\hat{G}^A]_{n\bm k}\propto \tau^\text{tr}$ upon summation over states $|n\bm k\rangle$. When multiple sources of scatterings are present, $\tau^\text{tr}$ ($\tau$) would correspond to the \emph{total} transport (usual) scattering time. Assuming there is no correlation between different sources of scatterings, then each scattering contributes to $\tau^\text{tr}$ (and $\tau$) independently as dictated by Matthiessen's rule, i.e., $1/\tau^\text{tr}=\sum_i(1/\tau_i^\text{tr})$ where $i$ labels the source of scattering, and hence $\rho_{xx}=\sum_{i}\rho_{i}$.

One notes that the same structure appears in $\sigma_\text{AH}^\text{I}$ as well. However, unlike $\sigma_{xx}$, the interband coherence effects including the interband velocity component and the interband off-shell scattering process (i.e. virtual transitions to states off the Fermi surface) are important~\cite{nagareview}, such that in the leading order diagrams each velocity vertex can be either intraband (dressed) or interband (not dressed). This suggests us to group the terms in $\sigma_\text{AH}^\text{I}$ according to the characters of velocity vertices involved, which naturally leads to three groups (see Fig.\ref{fig1}(b-d))~\cite{diagrams}, i.e., diagrams with 1) two interband vertices; 2) one interband vertex and one intraband vertex; and 3) two intraband vertices. Within each group, we can then identify and analyze the leading order terms.

Following this scheme, for group 1, the leading term is the single bubble without any off-shell scattering lines. This combined with $\sigma_{xy}^\text{II}$ gives the intrinsic contribution. For group 2, the part of dressed intraband velocity vertex along with two Green function legs, according to our previous discussion, is $\propto \tau^\text{tr}\propto\sigma_{xx}\approx 1/\rho_{xx}$ in terms of its scattering dependence. Meanwhile, its counterpart at the other vertex is of zeroth order in $\tau$. For the remaining middle part as indicated in the dashed box in Fig. \ref{fig1}(c), there must be at least one off-shell scattering line which brings a small factor $\propto 1/\tau_i\propto\rho_{i}$ for a scattering of type $i$. The important point is that this group of terms share a common factor $1/\rho_{xx}$ from the vertex part. Hence summing over $i$ leads to a generic form of $\sum_i c_i\rho_{i}/\rho_{xx}$ with coefficients $c_i$ that do not depend on the disorder concentration. Similar analysis applies to group 3. The two vertex parts each brings a factor of $1/\rho_{xx}$. In the middle part there must be at least two off-shell scattering lines ($\sim
\rho_i\rho_j$) or one three-scattering line ($\sim\rho_i$)~\cite{ndepend}. Note that the latter does not occur for dynamic disorder scattering (such as phonon)~\cite{lyo1973,Yang2011,brun2001} and it corresponds to the conventional skew scattering.

Collecting all the leading order terms and expressing them in terms of the longitudinal partial resistivities, we obtain the result
\begin{equation}\label{sigAH}
-\sigma_\text{AH}=c+\sum_i c_i\rho_i/\rho_{xx}+(\sum_{ij}c_{ij}\rho_i\rho_j+\sum_{i\in\text{S}}\alpha_i\rho_i)/\rho_{xx}^2.
\end{equation}
The $c$ term here corresponds to the usual intrinsic contribution. Transferring to the Hall resistivity $\rho_\text{AH}\simeq -\sigma_\text{AH}\rho_{xx}^2$~\cite{conversion}, we thus obtain the general scaling relation in Eq.(\ref{rhoAH}).

According to our
experimental setup, if we specialize Eq.(\ref{rhoAH}) to two major scattering mechanisms: one static and one dynamic, which could be tuned experimentally into competition, then we have $\rho_\text{AH}=c\rho_{xx}^2+(c_0\rho_{xx0}+c_1\rho_{xxT})\rho_{xx}+(c_{00}\rho_{xx0}^2+c_{01}\rho_{xx0}\rho_{xxT}+c_{11}\rho_{xxT}^2+\alpha\rho_{xx0})$,
where $\rho_{xx0}$ is the residual resistivity due to static impurities at low temperatures and $\rho_{xxT}$($=\rho_{xx}-\rho_{xx0}$) is due to dynamic disorders (mainly phonons at higher temperatures). Since there are only two independent variables out of the three ($\rho_{xx},\rho_{xx0},\rho_{xxT}$), if we choose to write in term of the two partial resistivities $\rho_{xx0}$ and $\rho_{xxT}$,
\begin{equation}\label{rhoAH2}
\rho_\text{AH}=\alpha \rho_{xx0}+\beta_0 \rho_{xx0}^2+\gamma\rho_{xx0}\rho_{xxT}+\beta_1\rho_{xxT}^2.
\end{equation}
with $\beta_0=c+c_{0}+c_{00}$, $\gamma=2c+c_0+c_1+c_{01}$, and $\beta_1=c+c_1+c_{11}$. One notes that there are only four scaling parameters: $\alpha$, $\beta_0$, $\beta_1$, and $\gamma$, which can be extracted from experiment. Except for the skew scattering term $\alpha\rho_{xx0}$ (which is absent for dynamic disorders~\cite{lyo1973,Yang2011,TianYuan,brun2001}), the scaling is bilinear in the two partial resistivities, and can be viewed as a nice combination of simple scalings of each type of scattering plus a cross term due to competitions. Furthermore the coefficients $\beta_0$ and $\beta_1$ of the two quadratic terms contain not only the intrinsic Berry curvature contribution $c$ but also side-jump contributions due to each scattering. In the following, we shall see that this result not only justifies the previous phenomenological scaling in the low conductivity regime ~\cite{TianYuan}, but also nicely captures its strong deviations in the high conductivity regime.

\begin{figure*}
\scalebox{0.78}{\includegraphics*{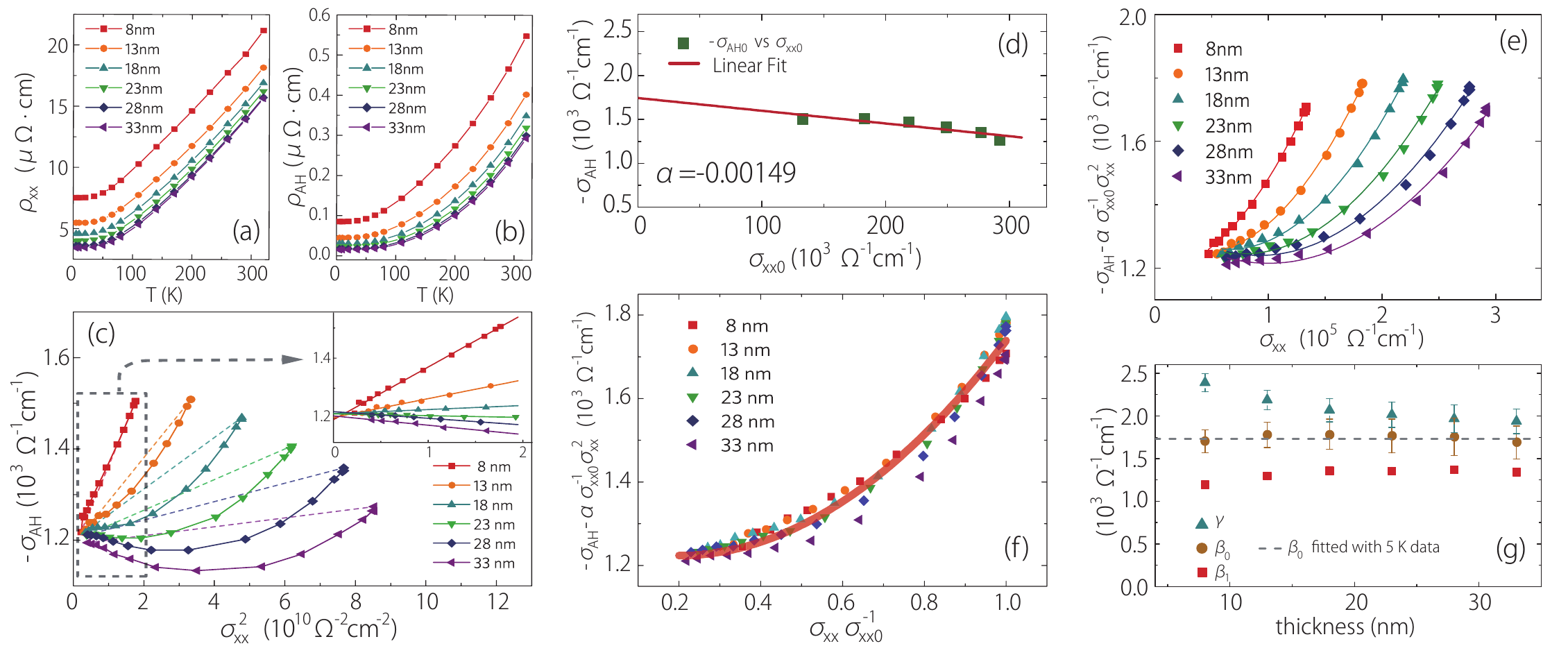}}
\caption{\label{fig2}(a,b) Temperature dependence of $\rho_{xx}$ and $\rho_{\rm{AH}}$ of epitaxial Fe films on MgO(001), respectively. (c) $\sigma_{\rm{AH}}$ versus $\sigma_{xx}^2$ for Fe films with various thicknesses. The inset shows the regime in which the data can be well fitted with the TYJ scaling, while systematic deviations from the TYJ scaling develop in the larger $\sigma_{xx}^2$ range. (d) $\sigma_{\rm{AH0}}$ versus $\sigma_{xx0}$ at 5 K. The red line is the linear fit and the slope $\alpha$ is the skew scattering coefficent. (e) Anomalous Hall conductivity with the skew scattering contribution subtracted $(-\sigma_{\textrm{AH}}-\alpha\sigma_{xx0}^{-1}\sigma_{xx}^2)$ versus $\sigma_{xx}$ for various film thicknesses. Fitting curves are based Eq.(\ref{sigscaling}). (f) $(-\sigma_{\textrm{AH}}-\alpha\sigma_{xx0}^{-1}\sigma_{xx}^2)$ versus $\sigma_{xx}\sigma_{xx0}^{-1}$ for different film thicknesses which almost all collapse onto a single curve. The red line is a guide of eye. (g) Fitting results for parameters $\beta_0$, $\beta_1$, and $\gamma$. The difference of the values of $\beta_0$ obtained from the fitting in present figure (brown dots) and from the interception in (d) is within the error bar.}
\label{fig1}
\end{figure*}

{\color{blue}{\it Experimental evidence.}}---Now we proceed to the discussion of our experimental data. Only after the previous work of Fe on GaAs(001)~\cite{TianYuan}, we find that the epitaxial Fe films on MgO(001) have much higher quality, as reflected by a much higher electrical conductivity for a film at the same thickness and temperature. It is from these high quality Fe film data that we discover significant deviations from the scaling proposed in \cite{TianYuan}, which call for a more rigorous theory for the scaling of AHE. In Fig.\ref{fig2}(a) and Fig.\ref{fig2}(b), the longitudinal resistivity and anomalous Hall resistivity measured from 5 K to 310 K are shown respectively, from which we can obtain the anomalous Hall conductivity $\sigma_{\textrm{AH}}$ and plot it against $\sigma_{xx}^2$ in Fig.\ref{fig2}(c). According to the scaling $\sigma_{\rm{AH}}=(\alpha\sigma_{xx0}^{-1}+\beta \sigma_{xx0}^{-2})\sigma_{xx}^2+b$ obtained phenomenologically in \cite{TianYuan} (hereafter referred to as the TYJ scaling), a linear dependence should have been expected, however, except for the 8 nm data, all the curves here show obvious deviation from the linear relation. Nevertheless, once zooming in the data in the low conductivity regime (where most of the data from Fe on GaAs located, see, Fig.4 in Ref.\cite{TianYuan}) as given in the inset, we can clearly see that they are indeed well described by the TYJ scaling. Comparing the present data with that of Fe on GaAs(001), we find that the breakdown of the TYJ scaling occurs in the high conductivity regime while in the real clean limit only the skew scattering term counts~\cite{nagareview}. In fact, by a closer look one could also find its traces in Fig.4 of Ref.\cite{TianYuan}, although not so obvious as here in Fig.\ref{fig2}(c).

Realizing the limitation of the scaling in Ref.\cite{TianYuan}, we now test whether our general scaling relation works here. For better comparison and analysis, rewriting Eq.(\ref{rhoAH2}) using conductivities $\sigma_{xx0}$ and $\sigma_{xx}$ gives
\begin{multline}\label{sigscaling}
-\sigma_{\rm{AH}}=\alpha\sigma_{xx0}^{-1}\sigma_{xx}^2
+(\beta_0+\beta_1-\gamma)(\sigma_{xx0}^{-1}\sigma_{xx})^2\\
+(\gamma-2\beta_1)\sigma_{xx0}^{-1}\sigma_{xx}+\beta_1.
\end{multline}
Compared with the TYJ scaling, one observes that the third term on the right hand side of Eq.\eqref{sigscaling} is formally new. We will expose the deeper connection between the two in a while.
To extract the four fitting parameters, first, by using the experimental data at 5 K in Fig.\ref{fig2}, we plot the residual conductivity $\sigma_{\textrm{AH0}}$ versus $\sigma_{xx0}$ for different film thicknesses in Fig.\ref{fig2}(d). Because of the limited thermal excitations at 5 K, $\sigma_{xx}\simeq\sigma_{xx0}$, Eq.(\ref{sigscaling}) becomes $-\sigma_{\rm{AH0}}=\alpha\sigma_{xx0}+\beta_0$ for different film thicknesses. Fitting the data using this relation (see Fig.\ref{fig2}(d)), we could first pin down the skew scattering coefficient $\alpha= -1.49\times10^{-3}$.

Then, by using the experimental data with variable temperature from 5 K to 300 K in Fig.\ref{fig2}, we show in Fig.\ref{fig2}(e) a series of $-\sigma_{\rm{AH}}$ \emph{vs} $\sigma_{xx}$ plot for different film thicknesses. Here we subtract the skew scattering contribution from $-\sigma_{\rm{AH}}$ in Eq.(\ref{sigscaling}), and use it to fit the corresponding experimental data, as shown by the solid lines in different colors in Fig.\ref{fig2}(e). On the other hand, one notes that after the subtraction of the skew term, the right hand side of Eq.(\ref{sigscaling}) is in fact a parabolic function of $\sigma_{xx0}^{-1}\sigma_{xx}$. Indeed, if plotting $(-\sigma_{\textrm{AH}}-\alpha\sigma_{xx0}^{-1}\sigma_{xx}^2)$ versus $\sigma_{xx}\sigma_{xx0}^{-1}$ using the experimental raw data for different film thicknesses in Fig.\ref{fig2}(f), we find strikingly that all the curves from Fig.\ref{fig2}(e) collapse on top of each other almost perfectly as indicated by the eye-guided line.

Plotting in Fig.\ref{fig2}(g) all the remaining three fitting parameters obtained in Fig.\ref{fig2}(e), we find that the three parameters are of the same order of magnitude and their values are quite stable across different film thicknesses. Particularly, $\beta_0$ and $\beta_1$ (which include intrinsic and side jump) are nearly constant. Here we have two separate ways to extract $\beta_0$: either from the interceptions of Fig.\ref{fig2}(d) using the experimental data at 5K or from the fitting of each curve in Fig.\ref{fig2}(e) using all the data from 5 K to 300 K. As shown in Fig.\ref{fig2}(g), the obtained values of $\beta_0$ agree with each other very well. $\gamma$ has a slight deviation for small thicknesses which could be originated from the finite size effect of dynamical scatterings when the film is very thin. Importantly, all three parameters quickly approach constant (bulk) values when the film thickness is greater than 20 nm.

\begin{figure}
\includegraphics[width=7.8cm]{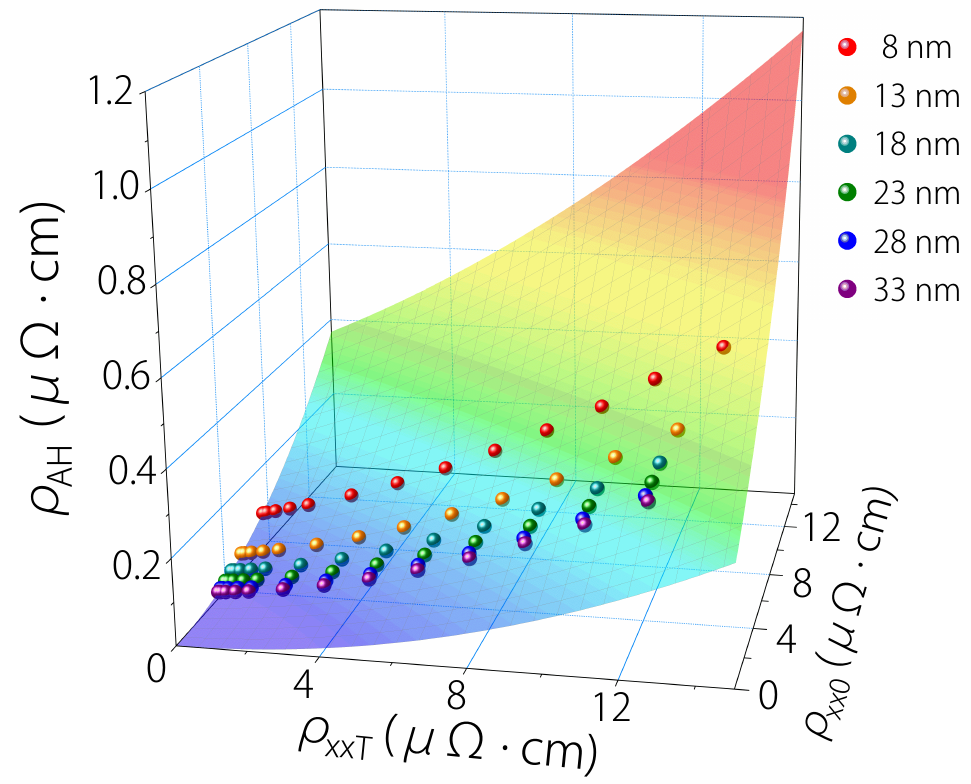}
\caption{The quadratic surface of $\rho_\text{AH}(\rho_{xx0},\rho_{xxT})$ according to Eq.\eqref{rhoAH2} with the values of four parameters determined in Fig.\ref{fig2}. The deviation of each raw data point from the surface is within the size of the dots.}
\label{fig3}
\end{figure}

Now with all four parameters $\alpha$, $\beta_0$, $\beta_1$, and $\gamma$ fitted from experiment, we plot in Fig.\ref{fig3} the surface $\rho_\text{AH}(\rho_{xx0},\rho_{xxT})$ as described by our general scaling Eq.\eqref{rhoAH2}. Remarkably, one observes that almost all the data points lie perfectly on this paraboloid surface, confirming the validity of our scaling relation.

{\color{blue}{\it Discussions.}}---We point out the important connection and contrast between our general scaling and the TYJ scaling in \cite{TianYuan}. We rearrange Eq.(\ref{sigscaling}) into the following form
\begin{multline}\label{expscaling}
-\sigma_{\rm{AH}}=\alpha\sigma_{xx0}^{-1}\sigma_{xx}^2+(\beta_0-\beta_1)\sigma_{xx0}^{-2}\sigma_{xx}^2+\beta_1+\\ (\gamma-2\beta_1)(\sigma_{xx0}^{-1}\sigma_{xx}-\sigma_{xx0}^{-2}\sigma_{xx}^2).
\end{multline}
While the first three terms on the right hand side have the same form as the TYJ scaling, one notes that the last term is negligible in both the high temperature limit (where $\sigma_{xx0}\gg \sigma_{xx}$) and the low temperature limit (where $\sigma_{xx}\simeq\sigma_{xx0}$). This hence explains the success of the phenomenological scaling as for Fe films on GaAs~\cite{TianYuan} as well as in the inset of Fig.\ref{fig2}(c), due to the smallness of the last term in this regime. Accordingly, for each film thickness in Fig.\ref{fig2}(c), by connecting with dash line the two data points at the highest and lowest temperatures respectively, we are in fact describing the corresponding AHE using the first three terms of Eq.\eqref{expscaling}. Obviously it significantly deviates from the reality where the real experimental points are connected by the eye-guided line, especially for thicker films with larger conductivity or relatively weaker static impurity scattring. Therefore, it is in the intermediate temperature regime where the static and dynamic impurity scatterings are competing with each other for the AHE, that the last term of Eq.\eqref{expscaling} manifests itself unambiguously as a critical factor to the AHE (one notes that the coefficient of the last term $(\gamma-2\beta_1)=c_0-c_1+c_{01}-2c_{11}$ is purely of scattering effect and is consistent with its interpretation as from the competition between scatterings). In view of the new experimental facts and deeper understanding presented in this work, the `proper scaling' found phenomenologically in the past \cite{TianYuan} is no longer appropriate.

Finally it should be noted that the coefficients of the second and the third term of Eq.\eqref{expscaling}: $\beta_0-\beta_1=c_0-c_1+c_{00}-c_{11}$ and $\beta_1=c+c_{1}+c_{11}$ also have new meaning compared with the scaling in \cite{TianYuan}. One observes that $(\beta_0-\beta_1)$ is also purely of scattering effects and the only presence of the intrinsic Berry curvature contribution $c$ in Eq.\eqref{expscaling} is in $\beta_1$. Our work demonstrates that with independent control of scatterings, it is possible to experimentally extract $\alpha_i$ and $(c+c_i+c_{ii})$ for each scattering source. Nevertheless, as in both Eq.\eqref{expscaling} and Eq.\eqref{rhoAH2}, the intrinsic contribution $c$ is entangled with the side-jump contributions $(c_i+c_{ii})$. Given that only the four parameters (and their linear combinations) can be extracted, therefore it seems challenging to unambiguously separate out the intrinsic part $c$ from such experimental approach. This will surely motivate further experimental studies in near future.

{\em Acknowledgements.} The authors thank D. L. Deng and Y. Shiomi for valuable discussions and thank Shan Guan for help with the figures.
This work was supported by MOST (Grant No. 2015CB921400, 2011CB921802), NSFC (Grants No. 11374057, No. 11434003 and No. 11421404), and SUTD-SRG-EPD2013062.


\begin{thebibliography}{15}
\bibitem [{nde()}]{ndepend}%
  \BibitemOpen
  \href@noop {} {\bibinfo  {journal} {Experimentally scattering effects are mostly
  controlled by tuning disorder concentrations. Here the
  relations are established regarding the quantities' dependence on the disorder
  concentration}\ }\BibitemShut {NoStop}%
\end{thebibliography}

%

\end{document}